%% file: Manuscript.tex
\documentclass[12pt]{article}
\usepackage{mathrsfs}
\usepackage{bm}
\usepackage{graphicx}
\usepackage{amssymb}
\usepackage{amsmath}
\usepackage{amsmath,amssymb,amsfonts, amsbsy, amsthm, booktabs, setspace,enumerate}
\usepackage{graphicx}
\usepackage[table]{xcolor}
\usepackage{tabularx}
\usepackage[utf8]{inputenc}
\usepackage[T1]{fontenc}
\usepackage{setspace}
\usepackage[round]{natbib}
 \usepackage{hyperref}
 \hypersetup{
     colorlinks=true,
     linkcolor=black,
     filecolor=black,
     citecolor = black, 
     urlcolor=blue,
     }

\usepackage{enumerate}
\usepackage{algorithm,algorithmic}
\usepackage[margin=1.0in]{geometry}
\usepackage{authblk}
\usepackage[notablist,nofiglist, nomarkers]{endfloat}

\newcommand{\bfm}[1]{\ensuremath{\mathbf{#1}}}

\def\bu{\bfm u}     \def\bU{\bfm U}     
     \def\bV{\bfm V}     
\def\bw{\bfm w}          
          
     \def\bY{\bfm Y}     
\def\bz{\bfm z}

\newcommand{\bfsym}[1]{\ensuremath{\boldsymbol{#1}}}

\def\bxi       {\bfsym {\xi}}

\renewcommand{\hat}{\widehat}

\graphicspath{{Figures/}}

\begin{document}

\title{Does Covid-19 Mass Testing Work? The Importance of Accounting for the Epidemic Dynamics}
\author[1]{Davide Ferrari}
\author[1]{Steven Stillman}
\author[1]{Mirco Tonin}
\affil[1]{Faculty of Economics and Management, Free University of Bolzano}
 \maketitle
\doublespacing
\begin{abstract}
Mass antigen testing has been proposed as a possible cost-effective tool to contain the Covid-19 pandemic. We test the impact of a voluntary mass testing campaign implemented in the Italian region of South Tyrol on the spread of the virus in the following months. We do so by using an innovative empirical approach which embeds a semi-parametric growth model - where Covid-19 transmission dynamics are allowed to vary across regions and to be impacted by the implementation of the mass testing campaign - into a synthetic control framework which creates an appropriate control group of other Italian regions. We find that the mass test campaign decreased the growth rate of Covid-19 by 39\% which corresponds to a reduction in the total additional cases of 18\%, 30\% and 56\% within ten, twenty and forty days from the intervention date, respectively. Our results suggest that mass testing campaigns are useful instruments for mitigating the pandemic.  
\end{abstract}

\noindent
{\bf Keywords}: {Mass testing; Covid-19; Growth model; Semi-parametric growth model; Synthetic controls.}

\newpage 

\input{Sec1.tex} 
\input{Sec1a.tex}  
\input{Sec2.tex} 
\input{Sec3.tex} 
\input{Sec4.tex} 

\newpage
 
\bibliographystyle{apalike}
\bibliography{biblio}

\end{document}

%% file: Sec1.tex
\section{Introduction}

A wide variety of interventions have been used in an attempt to stop the spread of Covid-19 across the globe (i.e. lockdown, quarantines, business closures, mobility restrictions, school closures) (\citealp{haug2020ranking}; \citealp{tian2020investigation}). Many of these are also the main candidate policies for stopping the spread of other large-scale epidemic outbreaks, such as Ebola. One particularly low-cost intervention that has been tried in a few locations is mass testing of a  population in order to identify asymptomatic carriers (\citealp{holt2020slovakia}; \citealp{atkeson2021economic}). Theoretical work suggests that mass testing could reduce daily infections by up to 30 percent (\citealp{Bosetti2020}), while \cite{Pavelkaeabf9648} evaluates the impact of a mass testing campaign undertaken in Slovakia in late 2020 and finds that it temporarily decreased the growth of Covid-19 by 70 percent. Importantly, many countries worldwide offer open public testing indicating that frequent mass testing is a feasible intervention (\citealp{Hasell2020}).

%

In this paper, we test the impact of a voluntary mass testing campaign implemented in the Italian region of South Tyrol on the spread of the virus in the following months. We do so by using an innovative empirical approach which embeds a semi-parametric growth model into a synthetic control framework. Specifically, we first use the synthetic control approach to create a control group which is a weighted-average of other Italian regions that best follow the dynamics of Covid-19 transmission in South Tyrol prior to the mass testing campaign (\citealp{abadie2010synthetic}; \citealp{abadie2003economic}).\footnote{We use this approach because it uses the data to derive the optimal control region for the South Tyrol. \cite{Mitze32293} and \cite{10.1093/ectj/utaa025} use this method to evaluate the impact of masks and other nonpharmaceutical interventions on the spread of Covid-19, however these papers implement it in a traditional framework which does not account for the underlying growth dynamics of an epidemic like Covid-19.} We then estimate on appropriately weighted data a semi-parametric growth model where Covid-19 transmission dynamics (i.e. growth rates) are allowed to vary across regions and to be impacted by the implementation of the mass testing campaign. Importantly, this approach is in the spirit of difference-in-difference models which compare changes in outcomes over time in a treated location to changes in outcomes over time in otherwise similar control locations and hence can isolate the impact of the mass testing campaign from national level policies regarding freedom of movement, business and school closures, hygienic measures, etc. since no other policy changed at the same time in South Tyrol. 

This approach has a number of benefits over what has so far been done in the literature that attempts to estimate the impact of public health interventions on the spread of Covid-19, as well as on other contagious diseases. Specifically, previous papers use either full parametric models (\citealp{wu2020generalized}; \citealp{aviv2020generalized}; \citealp{chenangnon2020use}; \citealp{pelinovsky2020logistic}) or traditional synthetic control or difference-in-difference models (\citealp{hsiang2020effect}; \citealp{tian2020investigation}; \citealp{mangrum2020jue}; \citealp{alexander2020mass}; \citealp{dave2021shelter}; \citealp{singh2021impacts}; \citealp{Mitze32293}; \citealp{10.1093/ectj/utaa025}). While fully parametric models based on biological growth theory have proved very useful for assessing the dynamics of disease outbreaks, their overly rigid a-priori parametric assumptions do not allow them to fit the typical variation in short- and medium-run dynamics seen in the spread of Covid-19. This is a potential concern with the findings of \cite{Pavelkaeabf9648}, especially since the testing campaign in Slovakia was a national level campaign and other policies were being implemented at the same time to reduce transmission rates. On the other hand, traditional difference-in-difference and synthetic control models, by assuming that treatment and control regions would follow parallel trends in a counterfactual reality without any intervention, are potentially biased because growth dynamics in different regions will depend on the prior contagion rate. 

Although there is a rich literature on estimating model-free growth curves in other fields, the use of non-parametric or semi-parametric approaches has not been sufficiently explored in the context of health policy analysis.\footnote{Exceptions include \cite{chenangnon2020use} which fits a flexible growth model curve to reported positive Covid-19 cases and \cite{lee2020estimation} which specifies dependence of the parameters in the Richards growth model on covariates to assess country-level risk factors due to Covid-19 within a hierarchical Bayesian framework. However, neither of these papers estimate the impact of a particular policy intervention.} We show that the estimated impact of the mass test campaign in South Tyrol is sensitive to how one models Covid-19 transmission dynamics and flexible models better fit the data and generate more robust estimates of the impact of the mass testing campaign. Furthermore, our approach has the added benefit that it generates estimates of the disease proliferation rate both with and without intervention, thus gaining insight on the transmission pathways.

Overall, we find that the mass test campaign in the South Tyrol decreased the growth rate of Covid-19 by 39\% (95\% confidence interval: 29-49\%). This corresponds to a reduction in the total additional cases  of 14\%, 18\%, 30\% and 56\% within seven, ten, twenty and forty days from the intervention date, respectively. Importantly, this large impact was achieved even though the campaign was entirely voluntary with no incentives to participate.\footnote{In Slovakia, individuals who did not participate were required to quarantine for ten days, hence it was not entirely a voluntary test.} Our results are  in line with the predictions made in \cite{Bosetti2020} based on an epidemiological model. We find a smaller impact than \cite{Pavelkaeabf9648} reports for the mass testing campaign in Slovakia; however, the Slovakian intervention featured a multiple round campaign in addition to concurrent interventions, including a one-week lockdown.\footnote{In line with our findings, a recent working paper by \cite{Kahanec2021Impact} uses a different approach to analyze the impact of the Slovakia campaign and finds that infections were reduced by about 25-30\% in the two weeks after individuals in some districts were tested a second time.}

%% file: Sec1a.tex
\section{The Mass Test} \label{sec:test}

The population of South Tyrol was invited to take part in a mass testing campaign in late November 2020 using rapid antigen tests, which involve a nasal and throat swab. Authorities set up around 300 testing centers, where professional health care workers carried out the tests, with the support of volunteers from the civil protection agency, the voluntary fire services and other organizations for handling the logistics and the administration. All residents were invited to participate, with the exception of children below the age of five, people with Covid-19 symptoms, those on sick leave, those who had tested positive and isolated in the last three months, and those who had recently tested positive or were in quarantine or self-isolating. People with a prior appointment for a PCR test, those regularly tested for work reasons, and individuals in social care were also not tested. 

Testing centers generally operated from 8am to 6pm from Friday, 20 November to Sunday, 22 November. During this period, people could show up at any of the centers throughout the region. In some municipalities, it was possible to register online and some published suggested centers and time slots based on the address of residence. It was also possible to be tested at some pharmacies and GPs in the period 18 to 25 November. People only needed a valid ID and a European Health Insurance card. They filled in a form with an email address, where they would receive, generally within a day, an encrypted file with the outcome, and a mobile number, where they would receive an SMS with the code to open the file. In case of a negative result, people were advised to continue following prevention measures like social distancing and mask wearing. In case of a positive result, people had to isolate for 10 days if asymptomatic and contact their doctor if they developed symptoms. 

Participation in the mass testing was voluntary and encouraged by a massive communication campaign, providing information (with material available also in Albanian, Arabic, English, French and Urdu, as well as in simple language for kids), as well as endorsements by public figures. The goal was to identify asymptomatic cases in the population and hence reduce virus transmission. The headline of the campaign was ``Together against coronavirus'', using appeals like ``Let's break the infection wave together and pave the way towards a gradual return to normality!''. In the end, 72 percent of eligible residents volunteered to be tested (\citealp{stillman2021communities}). In comparison, 83 percent of the eligible population in Slovakia decided to be tested when the alternative was to quarantine for 10 days. 

%% file: Sec2.tex
\section{Methods}

In this section, we first discuss how we create the optimal control group for evaluating the mass testing campaign in the South Tyrol. We then discuss how we implement the semi-parametric growth model, how it compares to parametric models and how we imbed a difference-in-difference type framework in this model. Finally, we discuss statistical inference.

 \subsection{Defining the control group} \label{sec:sc}

As there is no obvious a priori control group, in order to evaluate the effect of the screening intervention in South Tyrol, we constructed an optimal control group (Synthetic South Tyrol) using the synthetic control methodology of \cite{doi:10.1198/jasa.2009.ap08746}. This method constructs a weighted combination of data from the control group to approximate the behavior of the intervention group in terms of pre-intervention characteristics. 

Suppose we observe new cases $Y_{it}$ in regions $i= 1,\dots, N+1$ at times $t= 1, \dots,T$ and assume intervention occurs at time $T_0+1$ so that $1,\dots, T_0$ are pre-intervention periods. Without loss of generality the first region ($i=1$) represents South Tyrol, which is  exposed  to  the  intervention,  so  we  have $N$ remaining donor regions contributing to the synthetic control (in our case, the other 20 Italian regions or autonomous provinces in the donor pool). Let $\bw = (w_1, \dots, w_N)^\top$ be a $(N\times 1)$ vector of non-negative weights such that $\sum_{i=1}^N  w_i= 1$ and $w_i \ge 0$ for all $i=1, \dots, N$, where $w_i$ denotes the weight of region $i$ in the synthetic South Tyrol. Let $\bu_i=(x_{i1}, \dots, x_{iT_0}, z_{i1}, \dots, z_{iT_0} )^\top$ be a $(2T_0\times 1)$ vector containing pre-treatment values for cumulative number of cases ($x_{it}$) and number of tests ($z_{it}$) for region $i$. Thus, the vector $\bu_1$ contains pre-treatment values for South Tyrol and the matrix $(2T_0 \times N)$ matrix $\bU_0=(\bu_2, \dots, \bu_{N+1})$ with columns  $\bu_2, \dots, \bu_{N+1}$ contains pretreatment values for all the remaining donor regions.

The synthetic control method selects the vector of weights $\hat \bw$ that determines the best control region by solving the following quadratic optimization problem:
\begin{equation}\label{eq:prob1}
\text{min}_{\bw} \ \ (\bu_1 - \bU_0 \bw)^\top \bV (\bu_1 - \bU_0 \bw),  \ \ \text{subject to }\sum_{i=1}^N w_i =1,
\end{equation}
where $\bV$ is  a $2T_0 \times 2T_0$  symmetric  and  positive  semidefinite  matrix to  allow  different  weights  to  the  variables  in $\bU_0$ depending on  their  predictive  power  on  the  outcome. To stress dependence   on $\bV$, we use the notation $\hat \bw = \hat \bw(\bV)$ for the solution to (\ref{eq:prob1}). To select $\bV$, we minimize the mean squared prediction error (MSPE) of the outcome variable over pre-intervention periods as described in\cite{10.1257/000282803321455188} and \cite{doi:10.1198/jasa.2009.ap08746}. Specifically, if $\bY_1$ is the $(T_0 \times 1)$ vector with the values of the outcome variable (new cases) for South Tyrol and $\bY_0$ is the $(T_0 \times N)$ analogous matrix for the control units, then $\bV$ is selected by minimizing the $\text{MSPE}(\bV)= \Vert \bY_1 - \bY_0 \hat \bw(\bV) \Vert^2
$
over all positive definite and diagonal matrices $\bV$. 

The above procedure is implemented in the function \texttt{synth} of the R package \texttt{Synth} which uses the Nelder-Mead and BFGS algorithms as default options, and then picks the solution with the lowest MSPE \citep{JSSv042i13}. We then use the weights generated by this procedure when estimating the model described in the next section.
 
\subsection{Semi-parametric growth model} \label{Sec:models}

To analyze the dynamics of the Covid-19 epidemic over time, we develop a flexible semi-parametric growth curve approach. Let $x_t$ denote the cumulative size of the detected infected population at time $t$. In classic parametric growth curve analysis, the dynamics of $x_t$ is represented by the derivative of $x_t$, which is often assumed to have the form
\begin{equation} \label{eq:model1}
\dfrac{\partial x_t}{\partial t} = \rho \times x^p \times g(x_t),
\end{equation}
where $\rho$ is the intrinsic growth rate determining the time scale of the epidemic process, $p \in [0,1]$ is the ``deceleration of growth" parameter capturing different growth profiles and $g$ is a smooth non-increasing function of $x_t$ possibly depending on other model parameters. 

A number of well-known growth models can be recovered from Equation (\ref{eq:model1}) depending on the choice of $g$; although the literature is too vast to be covered here, we provide a few examples. A basic model is the exponential growth (EG) model corresponding to $g(x_t)= 1$. The EG model assumes that an epidemic continues to grow following the same process as in the past with the growth path completely specified by $p$: constant incidence ($p=0$), sub-exponential  growth  ($0<p<1$)  and  exponential growth ($p=1$); see, e.g. \cite{wu2020generalized}. Although this may be useful for representing the early stages of an epidemic, it is an upper bound scenario since outbreaks often slow down and reach saturation capacity after initial exponential growth. One popular extension is the generalized logistic growth (GLG) model   
$ g(x_t) =   x_t^p (1 - x_t/k)$,  where  $k$ represents the total size of the epidemic, i.e. the asymptotic number of infections over the whole epidemics \citep{wu2020generalized}. A slightly more flexible model is the generalized Richards growth (GRG) model  $ g(x_t) =   x_t^p [ 1 - (x_t/k)^a]$, where the parameter $a$ allows for deviations from the S-shaped dynamics of the classical GLG \citep{wu2020generalized, chowell2017fitting}. 

Despite the wide range of parametric growth models available, none of these rigid apriori specifications do a good job at fitting the rich dynamics observed in real Covid-19 data, which potentially leads to biased inferences. Instead, we estimate a more flexible model, referred to as a semi-parametric growth (SPG) model, that includes additional flexibility in two dimensions. 

We first add flexibility by estimating the function $g$ in the following semi-parametric specification:
\begin{equation} \label{eq:g}
 g(x_t) =  \exp\left\{ h(x_t) \right\} =  \exp\left\{ \sum_{j=1}^q \eta_{j} b_j(x_t)\right\},
 \end{equation}
where the $\eta_i$s are unknown coefficients and the $b_j(x_t)$ are given basis functions, such basis spline functions. The basis expansion $h(x_t) = \sum_{i=1}^q \eta_{i} b_j(x_t)\}$ allows us to approximate an arbitrary smooth function, provided that $q$ is large enough. Overall, this specification requires fewer assumptions about the data and fits the data better in situations where the true outbreak trajectory is hard to specify in advance, as is the case for the Italian Covid-19 data.

We also allow the growth rate parameter $\rho$ to depend on other explanatory variables. Since our main inferential interest is to assess the impact of the policy treatment on pre- and post-treatment growth differences, we assume the growth rate follows a log-linear difference-in-differences type of specification:
\begin{equation}\label{eq:log_growth}
\log\{\rho_{it}(\theta)\} =  \alpha + \beta \ \texttt{Int}_t  + \gamma \ \texttt{Reg}_i + \delta  \ \texttt{Int}_t \times \texttt{Reg}_i + \bxi^\top \bz_{it},   
\end{equation}
where  $\texttt{Int}_t$ is a dummy variable for the policy intervention taking value 0 up to the intervention date and 1 afterwards, $\texttt{Reg}_i$ is a dummy variable taking value 1 for the geographical region exposed to treatment (South Tyrol) and 0 otherwise, and $\bz_{it}$ is a $q\times 1$ vector of additional controls (such as the difference in the number of diagnostic tests compared to the previous day, weekly seasonality effects, etc). The overall vector describing the growth parameters is denoted by $\theta= (\alpha, \beta, \gamma, \delta,  \bxi^\top)^\top$. 

Our main inferential interest is in the coefficient $\delta$ for the interaction between the intervention and the geographical area. This parameter measures the effect of the policy intervention and is calculated in a similar way as in a traditional difference-in-differences model by comparing the change over time in the outcome variable for the South Tyrol compared to the change over time for the control region (Synthetic South Tyrol). We also assess the impact of policy intervention by examining its impact on the transmission growth rate. From equation (\ref{eq:log_growth}), the relative  change in the growth rate can be computed as 
$\Delta \rho_i(\theta)  = [\rho_{i1}(\theta) - \rho_{i0}(\theta)]/\rho_{i0}(\theta) = \exp\left\{ \beta + \delta \ \texttt{Reg}_i \right\} - 1
$. Therefore, the relative change in the treatment and control groups are $\Delta \rho_1 = \exp\{ \beta + \delta \} - 1$ and $\Delta \rho_0 = \exp\{ \beta \} - 1$, respectively.

\subsection{Statistical inference} \label{sec:inference}

Let $\{Y_{it}\}$ denote new COVID-19 cases observed in region $i$ at time $t$. We follow \cite{chenangnon2020use} and estimate the model parameters by assigning to each $Y_{it}$ an appropriate statistical distribution with mean $E[Y_{it}] = \partial x_{it}/\partial t = \rho_{it}(\theta) x_{it}^p  g(x_{it})$, where $x_{it}$ is the cumulative number of cases observed in region $i$ at time $t$ and $\rho_{it}(\theta)$ is a region- and time- specific growth rate parameter. For $Y_{it}$ we consider both the Poisson and negative binomial (NB) distributions, which are appropriate models for count data. Using a log-link function to relate the expected new cases to other predictors, the working statistical model for inference can be written as
\begin{align}\label{eq:stat_mod}
\log\left\{ E(Y_{it}) \right\} & = \log\left\{ \partial x_{it}/\partial t \right\}  = \log\{\rho_{it}(\theta)\} + p \log(x_{it}) + h(x_{it}),  
\end{align}
$i=0,1$ and $t=1,\dots, T$, where $\rho_{it}(\theta)$ is the time- and region-specific growth rate defined in Equation \ref{eq:log_growth}. 

Equation \ref{eq:stat_mod} represents a generalized additive model (GAM); e.g., see \cite{hastie1990generalized, wood2017generalized} for an introduction on GAMs. Estimates for the growth rate parameters $\hat \theta= (\hat \alpha, \hat \beta, \hat \gamma, \hat \delta, {\hat \bxi}^\top)^\top$, growth deceleration $\hat p$ and for the smooth function $\hat h(\cdot)$ are obtained by running a penalized likelihood estimator with penalty depending on a smoothing parameter for the non-parametric part of the model. We use thin-plate radial basis spline functions for the terms $b_i(x_{t})$  and smoothing parameter tuned via maximum likelihood estimation. 

Estimates are obtained using the function \texttt{gam} in the R package \texttt{mgcv} \citep{wood2015package}. See \cite{wood2017generalized} and references therein for details on the estimation procedure and implementation. Approximate variances and covariances for the estimated parameters are extracted using the function \texttt{vcov.gam} in the R package \texttt{mgcv}. An estimate of $\widehat{\Delta \rho_i}$, the relative growth change due to intervention as described in Section \ref{Sec:models}, is obtained by plugging-in the parameter estimates $\hat \beta$ and $\hat \delta$ into the expression $\Delta \rho_i(\theta)$. Since the parameter estimates are asymptotically normal, the standard errors for $\hat \Delta \rho_i$ can be derived using the Delta method (e.g., see \cite{van2000asymptotic}), obtaining 
$SE(\widehat{\Delta \rho_0}) = \exp\{ \hat \beta \} \times SE(\hat \beta)$ and  
$SE(\widehat{\Delta \rho_1}) = \exp\{ \hat \beta + \hat \delta\} \times \sqrt{SE(\hat \beta)^2 + SE(\hat \delta)^2 + \widehat{Cov}(\hat \beta, \hat \delta)}$, where $SE(\hat \beta)$ and $SE(\hat \delta)$ are the standard errors for $\hat \beta$ and $\hat \delta$ and $\widehat{Cov}(\hat \beta, \hat \delta)$ is an estimate of the covariance between $\hat \beta$ and $\hat \delta$. 

%% file: Sec3.tex
\section{Data and Results}

\subsection{Data}

The analysis in this paper focuses on the second COVID-19 outbreak wave that occurred in Italy in late 2020. Daily data on cumulative cases and number of tests from 1 September 2020 to 31 December 2020 on 19 Italian regions and 2 autonomous provinces were obtained from the online repository of the Italian Civil Protection Department \url{https://github.com/pcm-dpc/COVID-19}.  The autonomous province of Bozen/Bolzano - South Tyrol is the intervention group of our study as it was the only Italian territory that implemented mass screening in this time period. Residents of South Tyrol were invited to take a COVID-19 antigen rapid tests from Friday, 20 November through Sunday, 22  November; see Section \ref{sec:test} for details. 361,781 out of 500,607 (72.3 percent) eligible residents volunteered to take a test. 

\subsection{Constructing the control group}

\begin{figure}[h]
\centering
\includegraphics[scale=0.7]{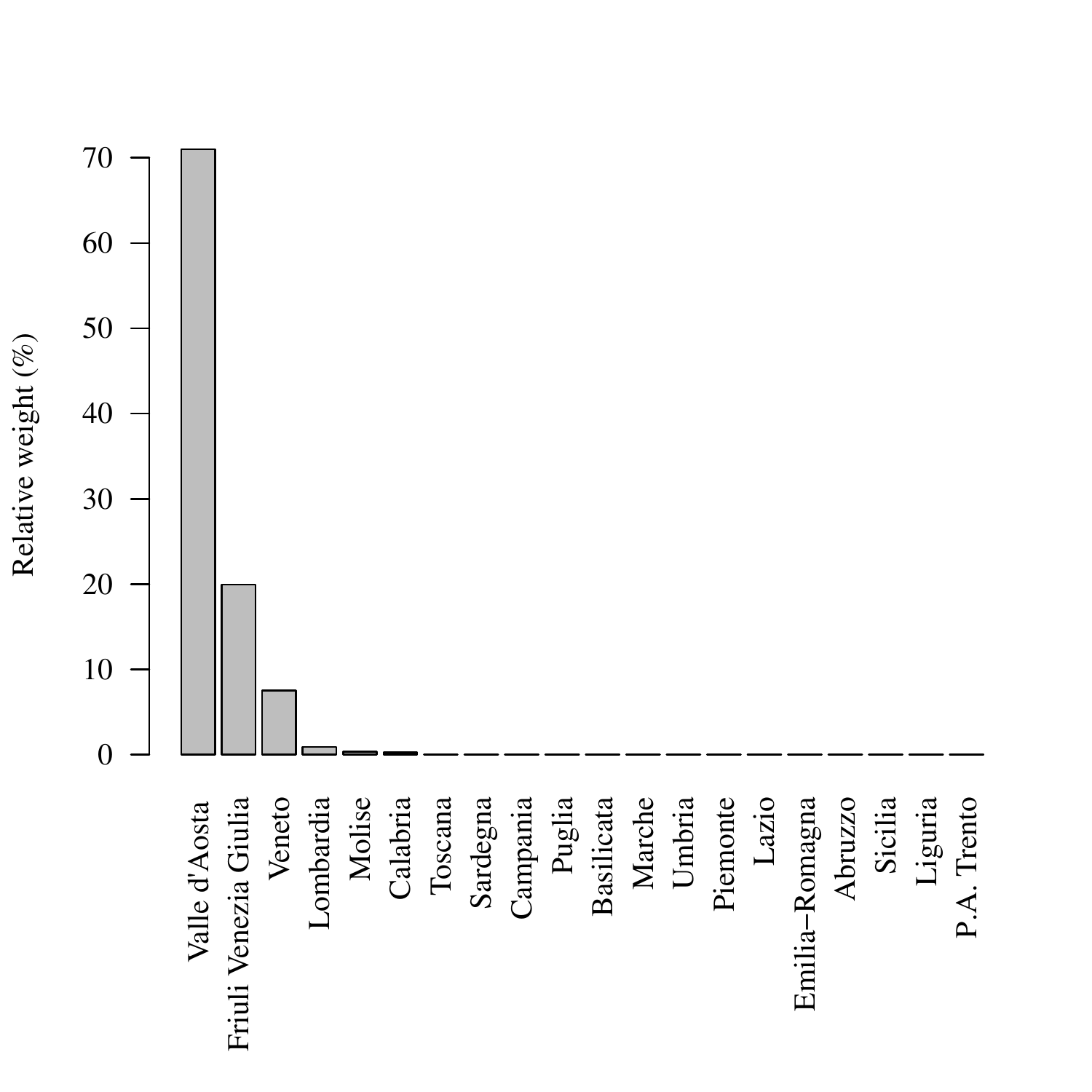} 
\caption{Synthetic control weights  used to construct the  synthetic control region.}
\label{fig:weights}
\end{figure}

The first step of our estimation process is to construct the control group of regions against which we will evaluate the impact of the mass testing in South Tyrol. Figure \ref{fig:weights} shows the estimated synthetic control weights for the 20 donor regions using the approach described in Section \ref{sec:test}. The regions Valle d'Aosta, Friuli Venezia Giulia and Veneto are by far the major contributors in the synthetic control with percent weights equal to 71.0\%, 20.0\% and 7.5\%, respectively. Valle d'Aosta is, like South Tyrol, a small mountain region in the North of Italy, but is not contiguous with South Tyrol. To assess the goodness-of-fit of this selection, we computed the R-squared type statistic  $R^2 = 1 -  \Vert \bY_1 - \bY_0 \hat \bw  \Vert^2/ \Vert \bY_1 - \bar{\bY} \Vert^2 = 0.871$, with $\bar{\bY}_1$ denoting the arithmetic average of the elements of vector $\bY_1$. Moreover, we found a  Pearson correlation coefficient between $\bY_1$ and $\bY_0 \hat \bw$ equal to $0.936$. Both the $R^2$ statistic and correlation coefficient show a very good match between South Tyrol and the synthetic control region in the pre-intervention period. 
 
\subsection{Results}

Table \ref{tab1} shows the estimates for the semiparametric growth models (SPGs) described in Section \ref{sec:inference} using Poisson (Pois) and negative binomial (NBin) response functions. Each model includes an indicator for being in the treatment region (\texttt{Reg}), an indicator for being in the time-period after the mass-testing intervention (\texttt{Int}), and interaction between the two, and an intrinsic growth parameter ($p$). We also estimate a second specification of each model that includes additional (unreported) controls (+Ctrl) that could impact short term (daily) movements of the outbreak trajectory, specifically i) difference in the number of new tests carried out, ii) day of the week fixed effects and iii) interactions between the number of new tests and day of the week fixed effects. 

Thus, overall we estimate four SPG models: Pois, Pois + Ctrl, NBin, and NBin + Ctrl. 
For each SPG model, we  report chi-squared statistics for the non-parametric component of our model corresponding to null hypothesis $H_0: h(x) = 0$. For comparison purposes, Table \ref{tab1} also shows in the lower part analogous setups for pure exponential growth models obtained by setting  $h(x_{it})=0$ in the semi-parametric models. For all the considered  models, we compute the adjusted R-squared and explained deviance statistics as well as the Akaike information criterion (AIC) and Bayesian information criterion (BIC) for model selection.

Based on both AIC and BIC model selection criteria, the best fitting model is NBin+Ctrl (negative binomial response with additional controls), which has an adjusted R-squared of 88.8\% and percent of explained deviance of 93.9\%. Overall, all the SPG models fit the data well with adjusted R-squared always exceeding 83\%. The standard EG models perform worse in terms of the goodness-of-fit metrics and for all the combinations of response distribution and predictors compared to the SPG models. The appropriateness of the semi-parametric models is also confirmed by the statistically significant chi-squared statistics for the non-parametric function $h$ in all considered cases. 

\begin{figure}[h]
\begin{tabular}{cc}
Pois & Pois + Ctrl \\
\includegraphics[scale=0.5]{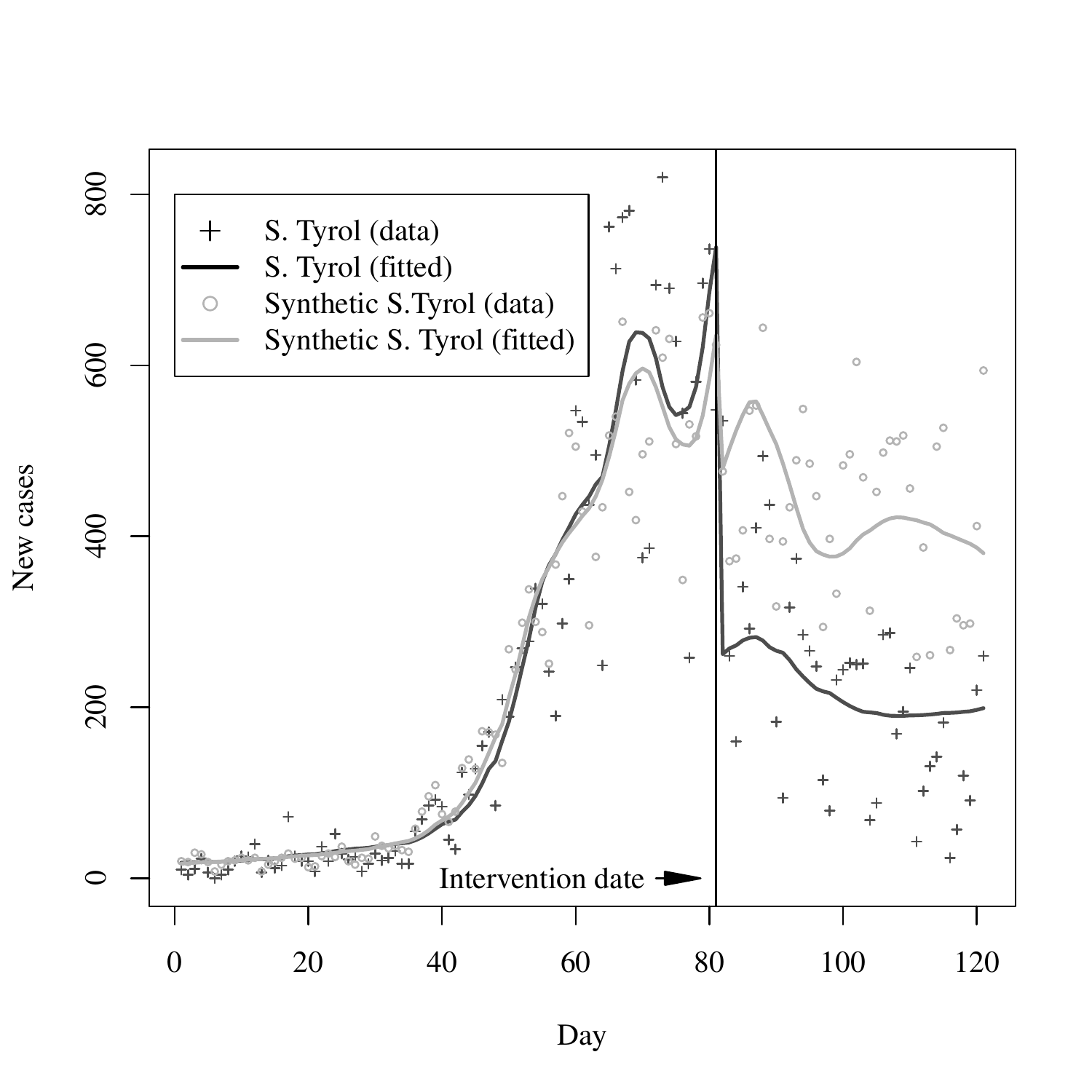} & \includegraphics[scale=0.5]{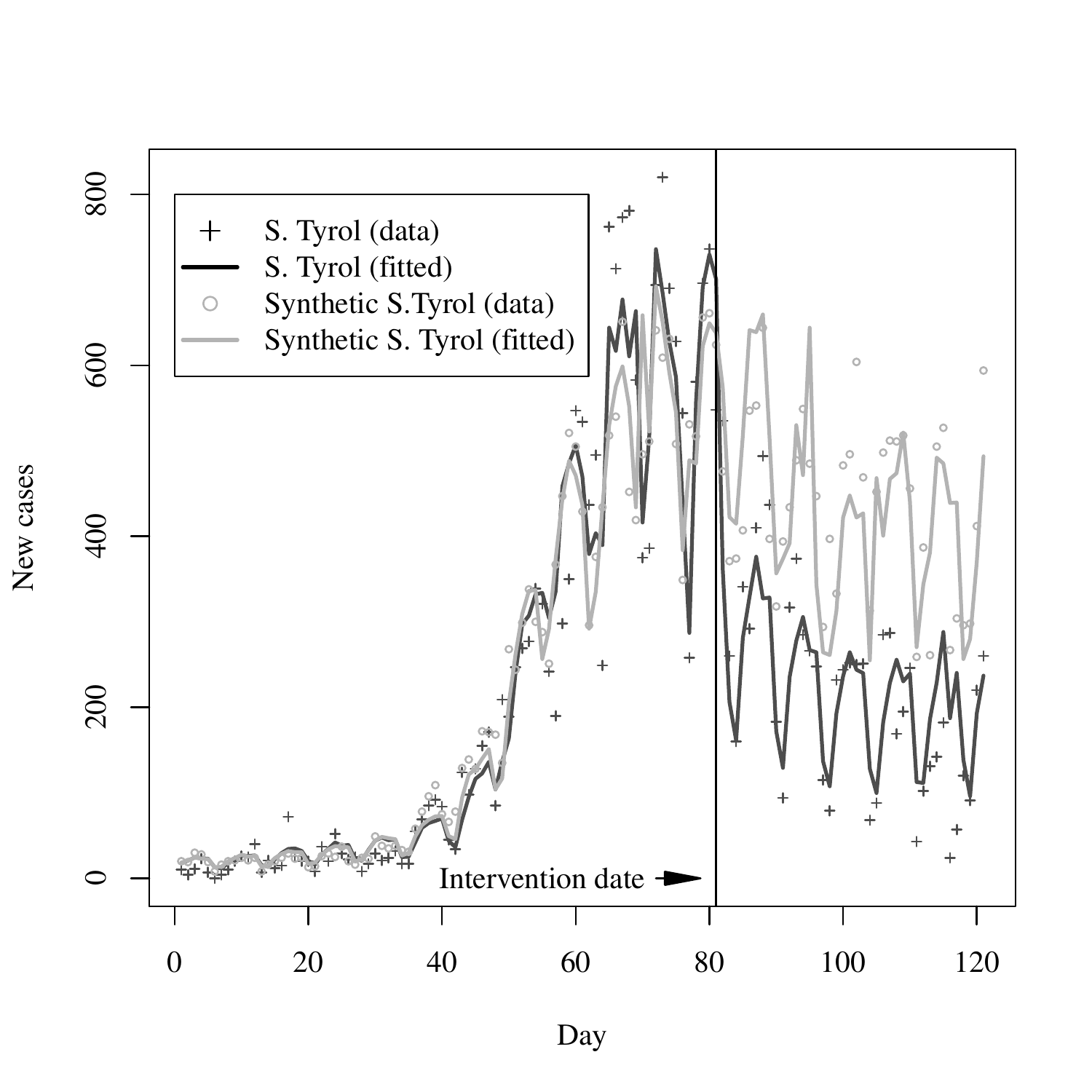} \\ 
NBin & NBin +Ctrl \\
\includegraphics[scale=0.5]{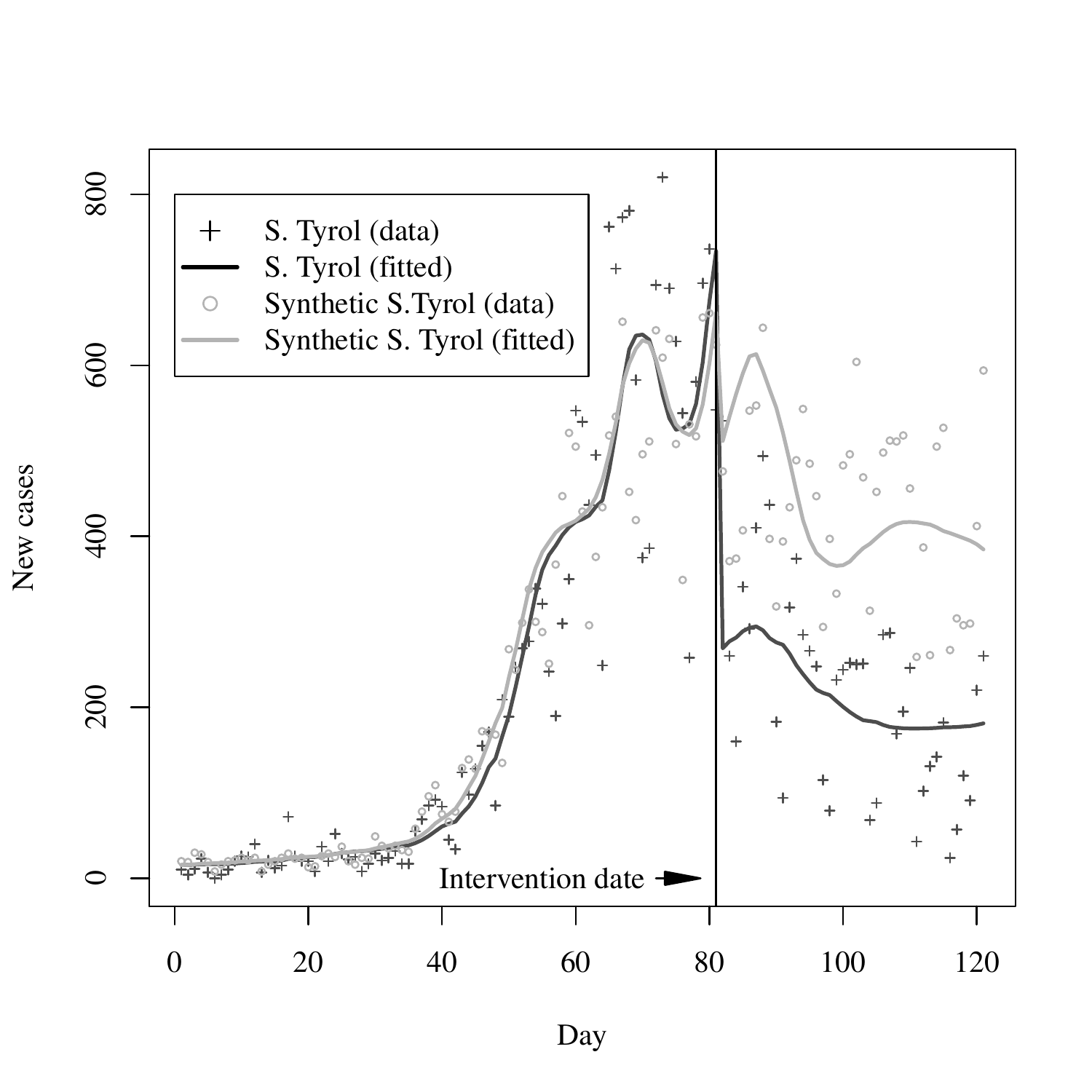} & \includegraphics[scale=0.5]{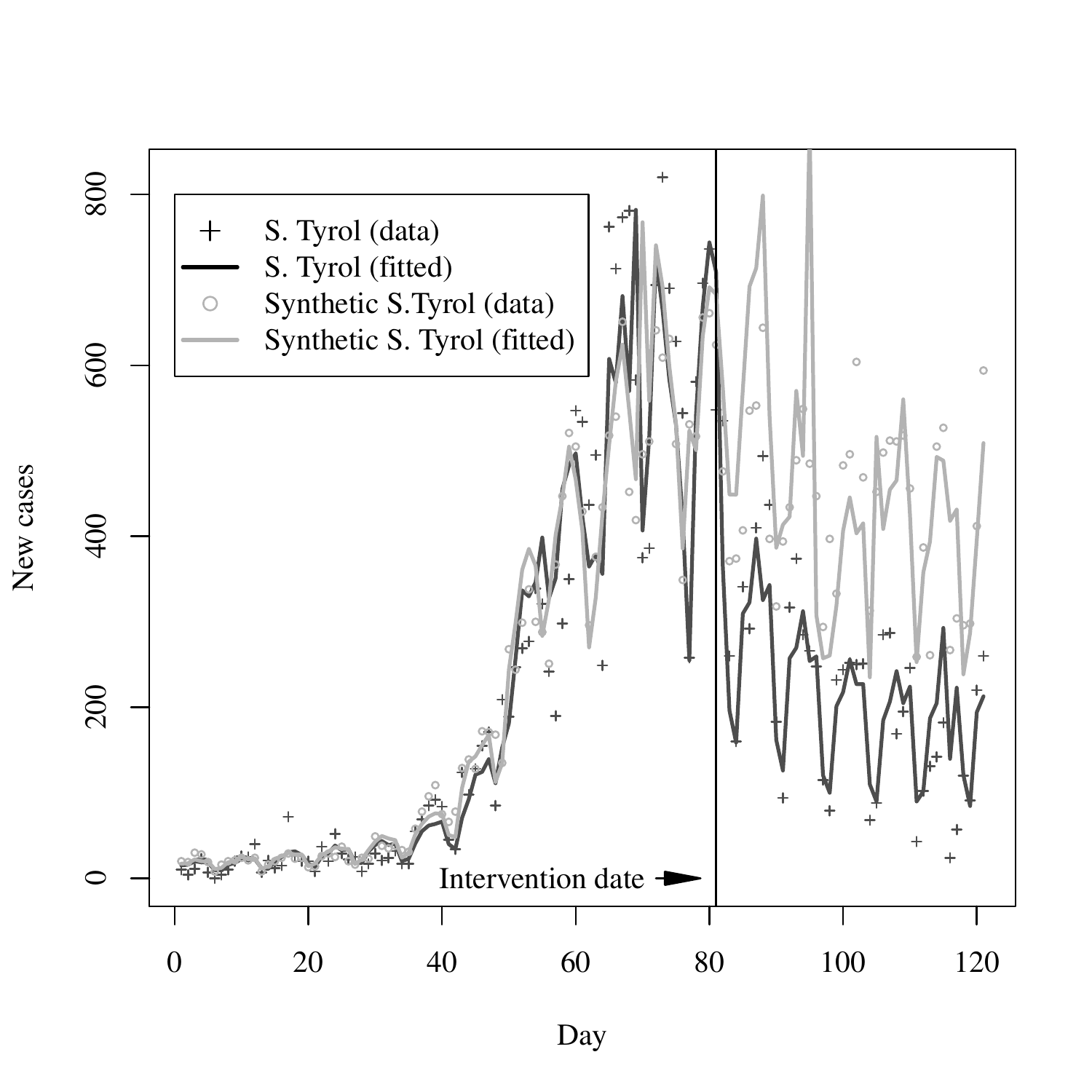} 
\end{tabular}
\caption{Fitted semi-parametric growth models against actual data on new cases for South Tyrol and synthetic control.The first and second row correspond, respectively, to Poisson (Pois) and negative binomial (NBin) models for new cases. The first column corresponds to models  without short-term control factors, while the second column shows models  with the additional controls included (+Ctrl).}
\label{fig1}
\end{figure}

Figure \ref{fig1} plots the estimated number of new cases $\hat y_{it}$ over time along with the actual observed data for both South Tyrol and the synthetic control region. The plots show the results from the Poisson and negative binomial models with and without the included additional short-term controls. Regardless of whether additional controls are included, the fitted models show a sizable and sudden decrease in the estimated new cases after intervention date. The models with additional covariates fit the short- and long-term data trajectories for both control and treatment regions remarkably well. 

In each of the SPG models, we find that the difference-in-differences interaction \texttt{Int$\times$Reg}, which indicates the impact of the mass testing campaign on Covid-19 growth in South Tyrol, is negative and highly significant. The estimated coefficient from our preferred SPG model NBin+Ctrl is -0.512 which implies a decrease of growth rate of $39\%$ (with 95\% confidence interval 29-49\%). These results are robust to both the response function used and whether controls are included with the estimated growth rate in these alternative specifications ranging from 38\% to 51\%. If instead one uses standard EG models, the estimated impacts are generally smaller, are less robust to model choice, and are less precisely estimated.

\begin{figure}[h]
\centering
\includegraphics[scale=0.9]{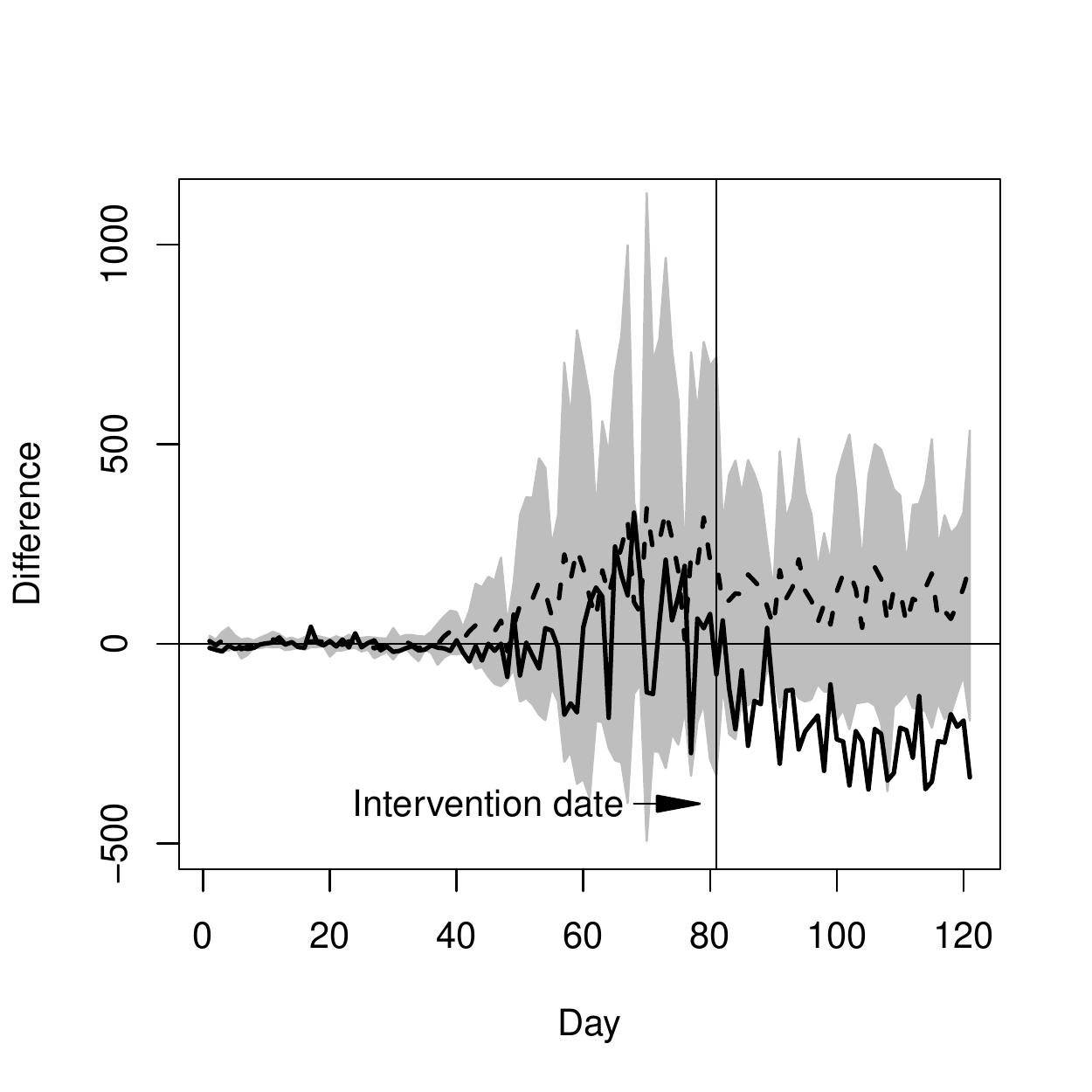} 
\caption{Differences between new cases in treatment and synthetic control regions over time, obtained by running the synthetic control method for each of the 21 Italian regions. The solid line corresponds to South Tyrol, the dashed line represents point-wise average difference for the remaining regions, the shaded area represents points within $1.96$ standard deviations from the point-wise average difference.}
\label{fig4}
\end{figure}

We next run a placebo test to gauge the possibility that our estimated impact of the mass testing campaign occurs by chance. We do this by matching each of the other 20 regions of Italy to synthetic versions of themselves and then estimating the SPG model (NBin + Ctrl) assuming there was an intervention at the same time as the real intervention in the South Tyrol. Figure \ref{fig4} compares the estimated impact in South Tyrol with the distribution of the estimates for the other regions. Further, we examine the distribution of the estimated interaction effect $\hat \delta$ for across placebo models. The estimated effect for the model with South Tyrol as treatment region is $\hat \delta =-0.512$ while a 95\% confidence interval based on the estimates for the other regions is $(-0.312, 0.160)$. The estimate for South Tyrol clearly stands out, with none of the other regions having large changes in transmission rates around the time of the intervention.

Finally, to evaluate the overall impact of the mass testing campaign, we compute the number of cases in South Tyrol over time by summing the predicted number of cases each day after the intervention assuming that the post-intervention transmission growth rate remained constant. Based on the estimates from our preferred model (SPG NBin+Ctrl), there were 1,743, 2,312, 5,754 and 10,533 fewer cumulative Covid-19 cases in South Tyrol compared to a scenario of unmitigated growth 7, 10, 20 and 40 days after the intervention, respectively. This corresponds to an overall reduction in cases of 14\%, 18\%, 30\% and 56\%). 

Our findings are in line with the predictions by \cite{Bosetti2020} which uses a SEIR dynamic ordinary differential equations model to predict outbreak dynamics under various contagion scenarios. When they consider the scenario where 75\% of the population is tested -- which is close to the 72\% observed in South Tyrol -- the predicted reduction in prevalence in the population within 10 days of mass testing is between 10\% and 30\% where we estimate a reduction of 14\% under the same time horizon.

\begin{table}[h]
\begin{tabular}{lrrrrrrr}
\hline
 & \multicolumn{4}{c}{Semi-parametric growth (SPG) models:}  \\
& \multicolumn{1}{c}{Pois} &  \multicolumn{1}{c}{Pois + Ctrl}  &  \multicolumn{1}{c}{NBin} &  \multicolumn{1}{c}{NBin + Ctrl} \\                     
\hline
&  \multicolumn{4}{c}{Parameter estimate (SE)}\\
\texttt{Int}       & -0.322 (0.034)* & -0.038 (0.035)\hspace{0.2cm} & -0.277 (0.278) \hspace{0.2cm}& -0.078 (0.201)\hspace{0.2cm} \\
\texttt{Reg}      &  0.072 (0.011)* & 0.085  (0.011)*  & 0.011 (0.063)\hspace{0.2cm} & 0.014 (0.047) \hspace{0.2cm}\\
\texttt{Int} $\times$ \texttt{Reg}  &  -0.754 (0.018)* &-0.569 (0.021)* &
-0.745 (0.118)* &-0.512 (0.094)*\\
$p$             &  0.189 (0.026)*    &  0.234 (0.027)* & 0.084 (0.047) \hspace{0.2cm} & 0.106 (  0.039)*\\
&  \multicolumn{4}{c}{$\chi^2$ statistic (approx df)*}\\
                   $ h(x)$             &   3413 (9)* &  2669 (9)* & 267.7 (9)* & 339.8 (9)*\\
                   & \multicolumn{4}{c}{Estimated growth change  $\widehat{\Delta \rho} \times 100 \%$ (SE)}  \\
South Tyrol        & -49.47 (0.47) & -38.38 (1.09) & -52.01 (4.82) & -39.22 (5.12) \\
Synth Control      &  7.43 (1.16) & 8.85 (1.21) & 1.13 (6.36) & 1.44 (4.78)  \\

                                       &  \multicolumn{4}{c}{Goodness-of-fit metrics}\\
 Adjusted-$R^2$ (\%) & 83.9   & 92.2 &  82.9    & 88.8\\
Dev. Explained (\%) & 88.9 & 94.5 & 88.1 & 93.9 \\
AIC                 & 7209.8 & 4436.1 & 2637.9  & 2507.7\\
BIC                 & 7258.7 & 4530.3 & 2689.3  & 2604.8\\
\hline
 & \multicolumn{4}{c}{Exponential growth (EG) models:}  \\
 & \multicolumn{1}{c}{Pois} &  \multicolumn{1}{c}{Pois + Ctrl}  &  \multicolumn{1}{c}{NBin} &  \multicolumn{1}{c}{NBin + Ctrl} \\                     
\hline
                   &  \multicolumn{4}{c}{Parameter estimate (SE)}\\
\texttt{Int}         &  -0.830 (0.013)*& -0.704 (0.013)* & -0.561 (0.128)*&  -0.426 (0.104)*\\
\texttt{Reg}      &   0.063 (0.011)*&  0.072 (0.011)*&-0.001 (0.088)\hspace{0.2cm} & 0.004 (0.070)\hspace{0.2cm} \\
\texttt{Int} $\times$ \texttt{Reg}  & -0.643 (0.017)* & -0.413 (0.018)* &-0.586 (0.151)* & -0.178 (0.126)\hspace{0.2cm} \\
$p$                  &   0.748 (0.005)* & 0.701 (0.005)*& 0.653 (0.025)*& 0.609 (0.023)*\\
                  & \multicolumn{4}{c}{Estimated growth change  $\widehat{\Delta \rho} \times 100 \%$ (SE)}  \\

South Tyrol        & -44.05 (0.73) & -28.87 (1.06) & -44.42 (6.79) & -15.96 (9.15) \\
Synth Control      &  6.46 (1.14) & 7.51 (1.19) & -0.15 (8.81) & 0.39 (7.05)  \\
                             &  \multicolumn{4}{c}{Goodness-of-fit metrics}\\
Adjusted-$R^2$ (\%) & 76.5 & 85.5 & 74.5& 56.8  \\
Dev. Explained (\%) & 82.2 & 89.3 & 72.5& 83.8 \\
AIC                 & 10560.7 &7013.6& 2830.4& 2726.5\\
BIC                 & 10578.2 &7076.4& 2851.3& 2792.8
\end{tabular}
\caption{Fitted semi-parametric growth (SPG) and exponential growth (EG) models. using Poisson (Pois) and negative binomial (NBin) response with additional control variables (+Ctrl) and without additional controls. For the parametric terms estimates are reported with standard errors in parenthesis. For the SPG models the $\chi^2$ statistics corresponding to null hypothesis ``$H_0: h(x) = 0$" are given with the approximate degrees of freedom (df). Significant results (p-value<0.01) are marked by ``*". For all the models we report  estimated  percent relative growth change  $\widehat{\Delta \rho} \times 100 \%$ for South Tyrol and synthetic control groups as well as goodness-of-fit statistics.}
\label{tab1}
\end{table}

%

%
%


%% file: Sec4.tex
\section{Conclusions}

In this paper, we employ an innovative empirical approach to study the impact of mass testing on Covid-19. In particular, we combine a synthetic control methodology to define a control group and a semi-parametric growth model to model the epidemiological developments. We show that using a semi-parametric approach rather than more standard parametric exponential growth models makes a difference in the evaluation of the impact of the mass testing that took place in the Italian region of South-Tyrol in November 2020 on Covid-19 growth. 

In our preferred specification, we find that mass testing reduced the growth rate of Covid-19 by 39\%. This suggests that mass testing can be an useful tool to contain the pandemic. Importantly, the mass testing campaign in South Tyrol was entirely voluntary and still managed to attract 72 percent of eligible residents to participate. This likely occurred because it was underlined that participating would help reduce the spread of Covid-19 and it could potentially lead to fewer restrictions in other dimensions. In many countries, public testing is openly available, suggesting that supply constraints are less and less binding. We find that a more organized approach to testing can make a big difference in reducing the spread of Covid-19. 